# Spectrum Sensing Based on Deep Learning Classification for Cognitive Radios

Shilian Zheng, Shichuan Chen, Peihan Qi, Huaji Zhou, and Xiaoniu Yang

*Abstract*—**Spectrum sensing is a key technology for cognitive radios. We present spectrum sensing as a classification problem and propose a sensing method based on deep learning classification. We normalize the received signal power to overcome the effects of noise power uncertainty. We train the model with as many types of signals as possible as well as noise data to enable the trained network model to adapt to untrained new signals. We also use transfer learning strategies to improve the performance for real-world signals. Extensive experiments are conducted to evaluate the performance of this method. The simulation results show that the proposed method performs better than two traditional spectrum sensing methods, i.e., maximum-minimum eigenvalue ratio-based method and frequency domain entropy-based method. In addition, the experimental results of the new untrained signal types show that our method can adapt to the detection of these new signals. Furthermore, we train the real-world signal detection experiment results show that the detection performance can be further improved by transfer learning. Finally, experiments under colored noise show that our proposed method has superior detection performance under colored noise, while the traditional methods have a significant performance degradation, which further validate the superiority of our method.**

## I. INTRODUCTION

WITH the large-scale deployment of 5G networks [1], the rapid emergence of the Internet of Things [2], and the rapid growth of various emerging technologies, the demand for wireless spectrum will become more and more urgent. As an opportunistic spectrum usage technology, cognitive radio [3] will greatly improve the efficiency of spectrum usage and is of great significance for mitigating the current situation of scarcity of spectrum resources. Spectrum sensing [4][5] is a key prerequisite for achieving dynamic spectrum access in cognitive radio. A lot of research on spectrum sensing have been done and many spectrum sensing algorithms have been proposed. Most methods design the corresponding decision statistics to detect the signal by studying the different characteristics between the signals and the noise. Therefore, the design of the decision statistics is very important for spectrum sensing performance. However, the design of the statistics requires extensive analysis and relies on domain-specific knowledge. As an end-to-end machine learning method that can automatically learn features from data, deep learning [6][7] has been widely used in image processing [8][9], speech recognition [10], natural language processing [11], and radio signal classification [12][13][14]. In this paper, we use deep learning for spectrum sensing. By formulating spectrum sensing as a classification problem with two categories, a solution based on deep convolutional neural network (CNN) is proposed. The performance of this method and traditional spectrum sensing methods are compared by extensive simulation experiments.

### A. Related Work

Traditional spectrum sensing methods include energy detection [15], cyclostationary feature-based detection [16], eigenvalue-based detection [17], frequency domain entropy-based detection [18], and power spectral density split cancellation-based method [19]. Among them, energy detection has been widely used for its advantages of easy implementation, but it suffers from the noise uncertainty seriously [20]. cyclostationary feature-based detection is superior in low signal-to-noise (SNR) scenario, but the method has high computational complexity and needs prior information of the signal. The eigenvalue-based detection makes the decision by calculating the eigenvalues and is robust to noise power uncertainty. The frequency domain entropy-based method relies on the difference of the distributions of the noise and the signal in the frequency domain. Inspired by the scalar transformation, the power spectral density split cancellation-based method uses the ratio of each subband power to the full band power as a test statistic, and can overcomes the impact of noise uncertainty effectively. However, the monitoring frequency band must be sparse in this method.

With the development of machine learning technology, there are also many studies that use machine learning for spectrum sensing, but most of them consider cooperative spectrum sensing. For example, the authors in [21] proposed several cooperative sensing algorithms based on support vector machine (SVM), weighted K nearest neighbor, K-means clustering and Gaussian mixture model. They used the energy of the received signal as the feature vector. In [22], K-means clustering and SVM techniques were used, but low-dimensional probability vectors were used as feature vectors. In [23], a CNN-based collaborative sensing method was proposed, which

Shilian Zheng, Shichuan Chen, Huaji Zhou, and Xiaoniu Yang are with Science and Technology on Communication Information Security Control Laboratory, Jiaxing 314033, China. (Email: lianshizheng@126.com, yxn2117@126.com)

Peihan Qi is with State Key Laboratory of Integrated Service Networks, Xidian University, Xi'an 710071, China.

improves the sensing performance and reduces the computational complexity. In [24], the authors presented an overview of applying various kernel-based learning (KBL) methods to cooperative spectrum sensing-related issues in cognitive radio networks for pursuing the superior nonlinear and high-dimensional signal processing capabilities of KBL methods over their linear counterparts.

In terms of single-node spectrum sensing, machine learning applications are just getting started. In [25], the authors proposed an artificial neural network (ANN)-based sensing method that uses energy and likelihood ratio test statistic as input features. In [26], the author also used ANN for spectrum sensing, with energy and cyclostationary features as the input. Using the same cyclostationary features, the authors in [27] turned to the CNN architecture for spectrum sensing. All these studies extract features in advance and then use neural networks to classify them. Therefore, their performance will largely depend on the pros and cons of previously extracted features. The most relevant work with this paper is the work carried out in [28][29]. In [28], the authors used the stacked autoencoder to detect OFDM signals. In [29], the authors used deep neural networks such as CNN and RNN to detect radar signals in the 3.5 GHz band based on the spectrogram. Better performance than the traditional detection methods was obtained. However, these two studies are concerned with the detection of specific signals, not the spectrum sensing problem in the general sense.

In this paper, we consider the spectrum sensing problem in the general sense, rather than the detection of a specific signal. We model the spectrum sensing problem as a classification problem with two categories and solve it based on deep learning. In the training phase, we use as many kinds of signals as possible and noise data to train the network, with the expectation that the trained network model can adapt to various unknown signals. Our method normalizes the power of the received signal and is essentially a robust approach to noise power uncertainty. In addition, most traditional spectrum sensing methods assume that the noise is additive white Gaussian noise (AWGN) which does not necessarily consistent with the actual situation. Since deep learning can automatically learn noise characteristics from data, our proposed method is expected to have the ability to adapt to actual non-ideal noise. We will evaluate the performance of our proposed method through extensive experiments.

*B. Contributions and Structure of the Paper*

The contributions of this paper are mainly as follows:
- We present spectrum sensing as a classification problem and propose a spectrum sensing method based on deep learning classification. The method uses the power spectrum of the signal as the input of the CNN and uses various types of signal data and noise data to train the network. The decision is made according to the confidence of the noise class. Since the method uses deep learning to automatically learn feature of signals and noise from data, it does not require prior feature extraction. Therefore, it can adapt to various noise types and can detect various signals including untrained signals.
- We conduct extensive experiments to verify the performance of the proposed method, specifically:
  a) We compare this method with traditional spectrum sensing methods, i.e., maximum-minimum eigenvalue ratio-based method and frequency domain entropy-based method. The results show that the probability of detection of our method is better than that of the two traditional methods under the same probability of false alarm.
  b) We conduct experiments with untrained signal types. The results show that the proposed method can adapt to the detection of these unknown signals, which shows that the model obtained by our method has strong generalization ability.
  c) We conduct transfer learning experiments for real-world signal detection. The results show that based on the model trained with the simulation data, the model can be fine-tuned with the real-world signal samples and the performance can be further improved when dealing with real-world signals.
  d) We analyze the performance of the method under colored noise. The results show that the proposed deep learning-based method has superior detection performance under colored noise, while the traditional methods based on maximum-minimum eigenvalue ratio and frequency domain entropy have a significant performance degradation under colored noise. This further validate the superiority of our method.

The rest of the paper is organized as follows. In Section II, we discuss the spectrum sensing problem, which can be modeled as a two-category classification problem. In Section III we give a detailed description of the proposed method based on deep neural networks. The simulation results are given in Section IV, and conclusions are made in Section V.

## II. PROBLEM FORMULATION

Spectrum sensing can be expressed as a binary hypothesis test problem as follows:

$$H_1: r(n) = hs(n) + w(n),$$
$$H_0: r(n) = w(n), \tag{1}$$

where $r(n)$ represents the received signal, $s(n)$ represents the transmitted signal, $w(n)$ represents the noise, and $h$ represents the channel gain. If no signal is present, the received signal contains only noise, otherwise it also contains the transmitted signal. From the perspective of classification, the problem shown in (1) can be expressed as a classification problem with two categories. One of the categories is signal and the other is noise. We will use a deep learning approach to solve this classification problem. We usually use the probability of detection and the probability of false alarm to evaluate the performance of the spectrum sensing algorithm, which are

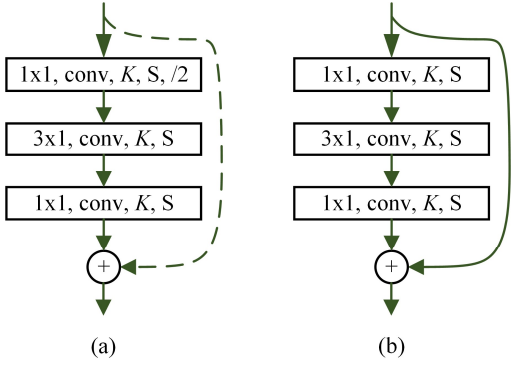

Fig. 1. Basic residual modules, where "conv" denotes the convolutional layer, "$K$" denotes the number of convolution kernels, "S" denotes padding to the same size, "/2" denotes downsampling by factor 2, and the dashed arrow denotes connection with downsampling. (a) Residual-block1 ($K$), (b) Residual-block2 ($K$).

TABLE I
THE RESIDUAL NETWORK STRUCTURE CONSTRUCTED

| Index | Layers | Output dimension |
|---|---|---|
| 1 | Input | 512x1 |
| 2 | 15x1, conv, 48 | 512x1 |
| 3 | 7x1, conv, 64 | 512x1 |
| 4 | 3x1, maxpool, /2 | 256x1 |
| 5 | Residual-block1 (64) | 128x1 |
| 6 | Residual-block2 (64) | 128x1 |
| 7 | Residual-block1 (128) | 64x1 |
| 8 | Residual-block2 (128) | 64x1 |
| 9 | Residual-block2 (128) | 64x1 |
| 10 | Residual-block1 (128) | 32x1 |
| 11 | 32x1, avgpool | 1x256 |
| 12 | fcx32 | 1x32 |
| 13 | Dropout (0.5) | 1x32 |
| 14 | Softmax, fcx2 | 1x2 |

defined as:

$$P_d = \Pr\{H_1|H_1\},$$
$$P_f = \Pr\{H_1|H_0\}. \tag{2}$$

## III. DEEP LEARNING-BASED SOLUTION

### A. The CNN Model

As a representative deep learning model, CNN [30] is widely used to solve problems such as image recognition and radio signal recognition. The CNN model considered in this paper is the residual deep learning model [31]. Compared with the traditional CNN, the residual network can maintain the performance improvement with the increase of the number of network layers without degradation. The basic residual unit considered in this paper is shown in Fig. 1. Based on these two basic modules, the details of the residual network structure constructed in this paper are shown in Table I, where "fc" represents the fully connected layer, "maxpool" represents max-pooling, and "avgpool" represents average pooling. First of all, we use the signal power spectrum as the network input. After the two basic convolutional layers and the max-pooling layer, six residual modules are cascaded, and the features activated are averaged through the average-pooling layer and sent to the fully connected layer with 32 neurons. In order to increase the generalization ability of the network, a dropout layer has been added. The final classification layer outputs the confidence vector of the signal class and the noise class. It should be noted that in order to deal with the problem of noise power uncertainty, we normalize the power of all received signals and calculate their power spectrum which serve as the network input.

### B. The Training Method

In order to accommodate the detection of various radio signals, our training data set includes as many types of wireless signals as possible. The purpose of this is to hope that the trained network model can also be generalized to the detection of other untrained signal types. In addition to the signal data, the training set also includes pure noise data of the same size.

The noise data can be AWGN noise or other colored noise. The training uses a basic stochastic gradient descent (SGD) method with momentum. The loss function is cross entropy function, i.e.,

$$l = \frac{1}{N}\sum_{i=1}^{N} o_{M(i)}, \tag{3}$$

where $N$ is the number of training samples, and the true label of the $i$-th sample is $M(i) \in \{0,1\}$, and $o_{M(i)}$ is the output confidence of the $i$-th sample in the true label dimension.

### C. The Decision Method

We judge the presence or absence of the signal based on the confidence of the output of the softmax layer. In general, the CNN classification often chooses the category with the highest confidence as the final classification result. If this strategy is adopted, for the two-category problem shown in this paper, the class with the confidence of greater than 0.5 will be selected as the detection result. However, because we can't control how much the model is biased towards one of the classes during training, it is difficult to control the detection performance if we use 0.5 as the decision threshold. We are improving this. Let $t_{\text{noise}}$ be the confidence of the noise class corresponding to the softmax output. The signal decision criteria we use are as follows:

$$H_1: 1 - t_{\text{noise}} > \gamma,$$
$$H_0: 1 - t_{\text{noise}} \leq \gamma, \tag{4}$$

where $\gamma$ is a certain threshold ($0 < \gamma < 1$). This threshold can be set according to the required probability of false alarm.

## IV. PERFORMANCE EVALUATION

### A. Data Generation

The signal data generated contains signals of 8 modulation types: BPSK, QPSK, 2FSK, 4FSK, 16QAM, 32QAM, 4PAM, and 8PAM. The pulse shaping filter uses a raised cosine filter. The normalized carrier frequency offset (relative to the sampling frequency) is randomly taken in the range [-0.1, 0.1]. The signal-to-noise ratio (SNR) ranges from -20 dB to 20 dB with an interval of 2 dB. Each signal sample contains 64 symbols, and the oversampling ratio is 8, so the length of each signal is 512. For each modulation type, 1000 samples at each signal-to-noise ratio were generated as training data and 500 samples were used as test data. The noise data in the training set is AWGN data or colored noise data generated by the simulation, and the length is also 512. The number of noise samples is the same as the signals. In order to verify the ability of the trained model to detect signals with other untrained types, we also simulated signal data of three modulations, 8PSK, 8FSK and 64QAM, as test samples for unknown signals.

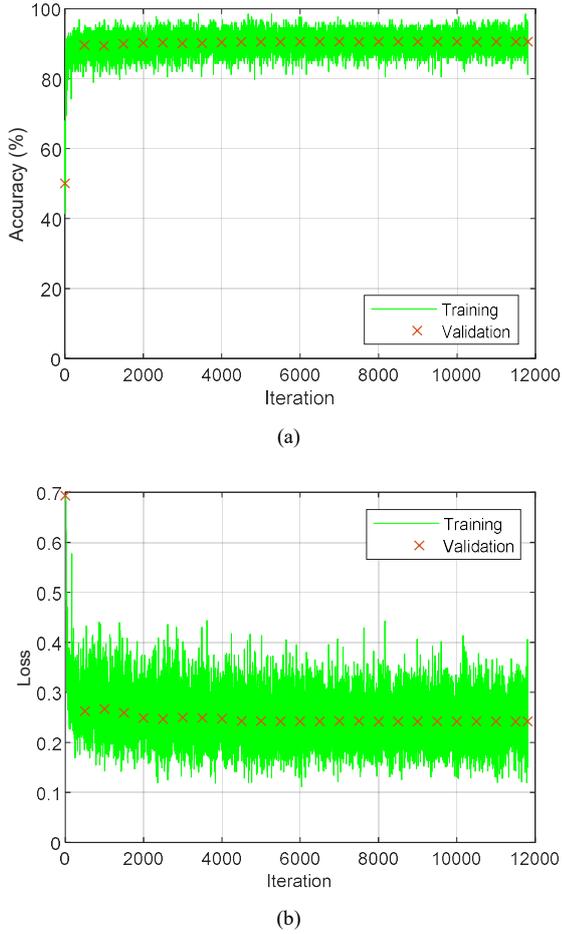

(a)

(b)

Fig. 2. The training process. (a) Accuracy and (b) loss.

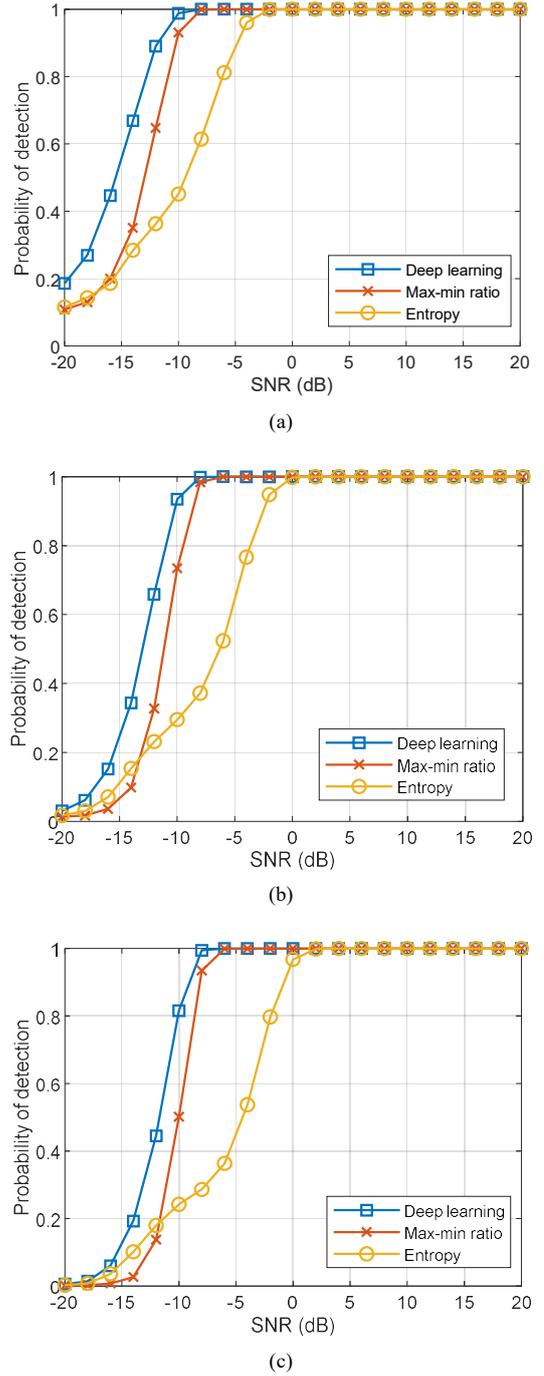

(a)

(b)

(c)

Fig. 3. Comparison with traditional methods. (a) $P_f = 0.1$, (b) $P_f = 0.01$, and (c) $P_f = 0.001$.

### B. Simulation Results

#### 1) Basic training process

Training was performed on the NVIDIA V100 GPU card. All parameters were randomly initialized with a Gaussian distribution. The mini-batch size was set to 128 during training, and SGD with momentum was used as the training method with a momentum factor of 0.9. The initial learning rate was 0.01. Every 3 epochs, the learning rate was reduced to 1/10 of the previous value. The network was trained with 9 epochs and verification was performed every 500 iterations. Fig. 2 shows the training process. It can be seen that the training process converges fast. After convergence, the classification accuracy rate on the entire test data set is 90.55%. Since the SNR of the entire test set is from -20 dB to 20 dB, the performance at very low SNR will lower the overall accuracy. We will give detailed detection accuracy under different SNRs in the next subsection.

#### 2) Comparison with traditional methods

We compare the performance of our method with the traditional spectrum sensing methods. The traditional spectrum sensing method used as comparison is maximum-minimum eigenvalue ratio-based method and frequency domain entropy-based method. Fig. 3 shows the experimental results. The noise order to verify the ability of the neural network model trained with the several signal types to detect other types of signals that have not been trained, we simulated 8PSK, 8FSK and 64QAM signals to evaluate the performance. Fig. 4 shows the results, where the probability of false alarm is set to 0.01. It can be seen that these signals can still be detected with high probability. The detection performance of our method is also superior to the traditional methods, which further illustrating the ability of our proposed method to adapt to sensing of various unknown signals.

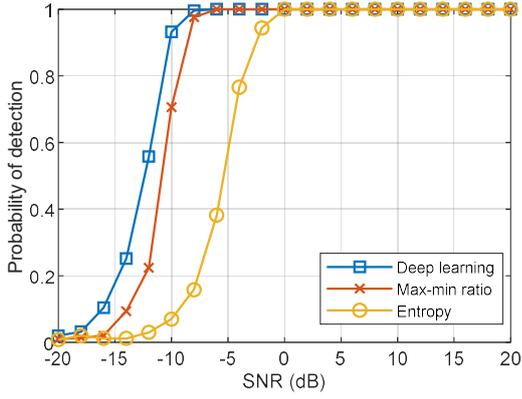

(a)

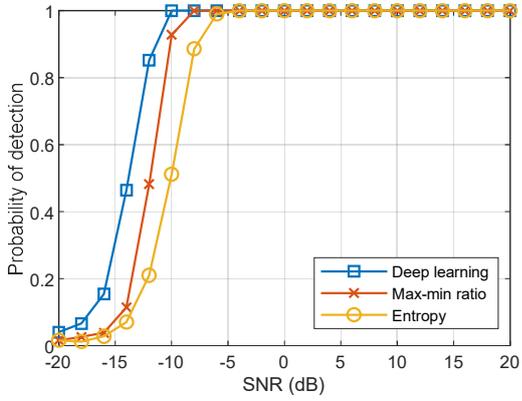

(b)

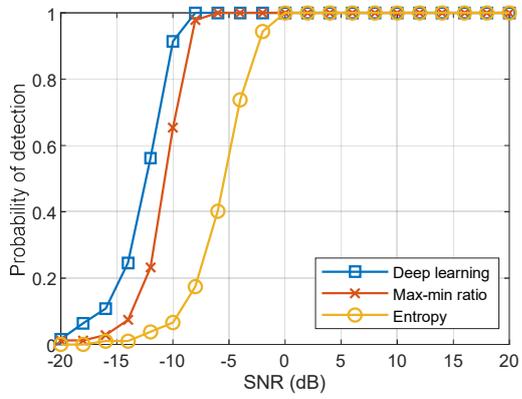

(c)

Fig. 4. Detection performance of new signals (untrained signal types), $P_f =$ 0.01. (a) 8PSK, (b) 8FSK, and (c) 64QAM.

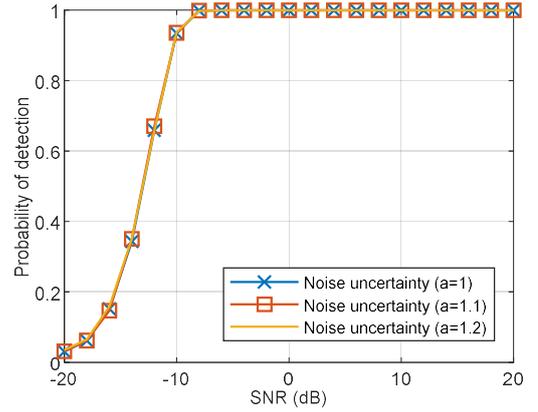

Fig. 5. Influence of noise power uncertainty.

*4) Influence of noise power uncertainty*

We use the signal power spectrum as the input to the neural network. Before calculating the power spectrum, we normalize the signal in the time domain to overcome the influence of noise power uncertainty. To further illustrate the robustness of our approach to noise power uncertainty, we performed simulation experiments. In the experiment, the noise power range is $[\frac{1}{a}P, aP]$, where $P$ is the real noise power and $a \geq 1$ is the noise uncertainty factor. The noise considered was AWGN. We take 0.01 as the target probability of false alarm. Simulations show that when there is noise power uncertainty, the false alarm probability remains unchanged. The detection probability is shown in Fig. 5. It can be seen that under different noise power uncertainty factors, the detection performance of the proposed method is hardly affected, which demonstrates its robustness to noise power uncertainty.

*5) Transfer learning for real-world signals*

In order to further verify the performance of the method in detection of real-world signals, we carried out experiments with the Aircraft Communications Addressing and Reporting System (ACARS) [32] signals collected over the air. We obtained 1000 ACARS samples, half of which were used as training set for transfer learning and the other half as test set. It should be noted that although the signal samples are captured over the air, the noise samples are still generated with simulation, not the noise in the actual environment. In the transfer learning (TL), the model obtained by training on the above simulation data was used as the basic model, and this model was fine-tuned by using the real-world ACARS signals with injected simulated AWGN noise. The learning rate was set simulated was AWGN. It can be seen that under various false alarm probabilities, the signal detection probability obtained by our method is superior to the traditional methods. Especially in the low SNR, the performance gain of our method is more obvious.

*3) Detection of new signals*

Because the deep learning-based spectrum sensing method constructed in this paper is a signal detection method in the general sense, rather than a specific signal detection method. In to 0.0001, and the transfer learning took 1 epoch. Fig. 6 shows the experimental results on the test set with the false alarm probability set to 0.01. It should be noted that the SNRs in the figure is obtained by injecting man-made noise to the signals. Since the ACARS signals collected over the air already contain a certain amount of noise, the actual SNR is lower than the value on the coordinate axis in the figure. It can be seen that the model without transfer learning performs better than max-min eigenvalue ratio-based method and has similar detection performance to the frequency domain entropy-based method. However, after transfer learning, the detection performance of our method has been further improved, which is better than the other methods.

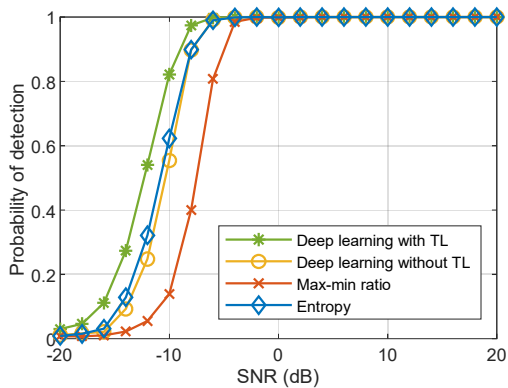

Fig. 6. Detection performance of real-world ACARS signals.

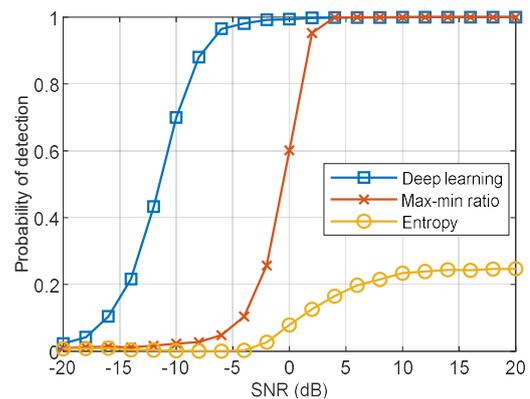

Fig. 7. Performance with pink noise. $P_f = 0.01$.

*6) Performance under colored noise*

Most traditional spectrum sensing methods are based on the assumption that the noise if white. However, the noise experienced in practice is not necessarily the ideal white noise. We evaluated the performance of the methods under colored noise. Fig. 7 shows the simulation results, in which the noise used is pink. It can be seen that our proposed deep learning-based method has superior performance in the case of pink noise, which is similar to the ideal white noise. However, the traditional max-min eigenvalue ratio-based method and frequency domain entropy-based method have significant performance degradation under pink noise. For the frequency domain entropy method, the detection probability is still low even with SNR up to 20 dB. This experiment verifies the ability of deep learning to automatically learn noise features from the data.

## V. Conclusions

We have considered spectrum sensing as a classification problem with two categories and proposed a spectrum sensing method based on deep learning. The simulation results show that the proposed method outperforms the traditional maximum-minimum eigenvalue ratio-based method and the frequency domain entropy-based method, which demonstrates the effectiveness of the proposed method. The method also has the ability of generalization and can adapt to detecting a variety of untrained signals. When dealing with real-world ACARS signals, the performance of our proposed method can be further improved by transfer learning. Since deep learning can automatically learn noise characteristics from the data, our method can still maintain the performance close to AWGN in the case of pink noise, while the performance of the traditional methods degrades significantly. Except for the real-world data used in the ACARS experiment, most of the experiments in this paper are based on simulation data. In the future, we will conduct experiments with large-scale real-world signal data collected over the air to further evaluate the performance of the method in the actual environment.

## References

[1] A. Gupta and R. K. Jha, "A survey of 5G network: Architecture and emerging technologies," *IEEE Access*, vol. 3, pp. 1206-1232, 2015.

[2] A. Al-Fuqaha, M. Guizani, M. Mohammadi, M. Aledhari, and M. Ayyash, "Internet of Things: A survey on enabling technologies, protocols, and applications," *IEEE Communications Surveys & Tutorials*, vol. 17, no. 4, pp. 2347-2376, 2015.

[3] S. Haykin, P. Setoodeh, "Cognitive radio networks: The spectrum supply chain paradigm," *IEEE Transactions on Cognitive Communications and Networking*, vol. 1, no. 1, pp. 3-28, 2015.

[4] T. Yucek and H. Arslan, "A survey of spectrum sensing algorithms for cognitive radio applications," *IEEE Communications Surveys & Tutorials*, vol. 11, no. 11, pp. 116-130, 2009.

[5] H. Sun, A. Nallanathan, C.-X. Wang, and Y. Chen, "Wideband spectrum sensing for cognitive radio networks: a survey," *IEEE Wireless Communications*, vol. 20, no. 2, pp. 74-81, 2013.

[6] Y. LeCun, Y. Bengio, and G. Hinton, "Deep learning," *Nature*, vol. 521, no. 7553, pp. 436-444, 2015.

[7] I. Goodfellow, Y. Bengio, and A. Courville, *Deep Learning*, MIT Press, 2016.

[8] W. Liu, Z. Wang, X. Liu, N. Zeng, Y. Liu, and F. E. Alsaadi, "A survey of deep neural network architectures and their applications," *Neurocomputing*, vol. 234, pp. 11-26, 2017.

[9] A. Ioannidou, E. Chatzilari, S. Nikolopoulos, and I. Kompatsiaris, "Deep learning advances in computer vision with 3d data: A survey," *ACM Computing Surveys (CSUR)*, vol. 50, no. 2, 2017, pp. 20.

[10] G. Hinton, L. Deng, D. Yu, A.-R. Mohamed, N. Jaitly, A. Senior, V. Vanhoucke, P. Nguyen, T. Sainath, G. Dahl, and B. Kingsbury, "Deep neural networks for acoustic modeling in speech recognition: The shared views of four research groups," *IEEE Signal Process Magazine*, vol. 29, no. 6, pp. 82–97, 2012.

[11] R. Socher, Y. Bengio, and C. D. Manning, "Deep learning for nlp (without magic)," in Tutorial *Abstracts of ACL 2012*, pp. 5, 2012.

[12] S. Zheng, S. Chen, L. Yang, J. Zhu, Z. Luo, J. Hu, and X. Yang, "Big data processing architecture for radio signals empowered by deep learning: Concept, experiment, applications and challenges," *IEEE Access*, vol. 6, pp. 55907-55922, 2018.

[13] S. Zheng, P. Qi, S. Chen, and X. Yang, "Fusion methods for CNN-based

automatic modulation classification," *IEEE Access*, vol. 7, pp. 66496-66504, 2019.

[14] S. Chen, S. Zheng, L. Yang, and X. Yang, "Deep learning for large-scale real-world ACARS and ADS-B radio signal classification," *IEEE Access*, 2019.

[15] F. F. Dighan, M.-S. Alouini, and M. K. Simon, "On the energy detection of unknown signals over fading channels," in *Proc. IEEE ICC*, 2003, vol. 5, pp. 3575-3579.

[16] J. Lundén, S. A. Kassam, and V. Koivunen, "Robust nonparametric cyclic correlation-based spectrum sensing for cognitive radio," *IEEE Transactions on Signal Processing*, vol. 58, no. 1, pp. 38-52, 2010.

[17] Y. Zeng and Y.-C. Liang, "Eigenvalue-based spectrum sensing algorithms for cognitive radio," *IEEE Transactions on Communications*, vol. 57, no. 6, pp. 1784-1793, 2009.

[18] Y. Zhang, Q. Zhang, and S. Wu, "Entropy-based robust spectrum sensing in cognitive radio," *IET Communications*, vol. 4, no. 4, pp. 428-436, 2010.

[19] P.-H. Qi, Z. Li, J.-B. Si, *et al.*, "A robust power spectrum split cancellation-based spectrum sensing method for cognitive radio systems," *Chinese Physics B*, vol. 23, no. 12, pp. 128401 1-11, 2014.

[20] E. H. Gismalla, "Performance Analysis of the Periodogram-Based Energy Detector in Fading Channels," *IEEE Trans. Signal Process.*, vol. 59, no. 8, pp. 3712-3721, Aug. 2011.

[21] K. M. Thilina, K. W. Choi, N. Saquib, and E. Hossain, "Machine learning techniques for cooperative spectrum sensing in cognitive radio networks," *IEEE Journal on Selected Areas in Communications*, vol. 31, pp. 2209–2221, November 2013.

[22] Y. Lu, P. Zhu, D. Wang, and M. Fattouche, "Machine learning techniques with probability vector for cooperative spectrum sensing in cognitive radio networks," in *2016 IEEE Wireless Communications and Networking Conference*, April 2016, pp. 1–6.

[23] W. Lee, M. Kim, D. Cho, and R. Schober, "Deep sensing: Cooperative spectrum sensing based on convolutional neural networks," *CoRR*, vol. abs/1705.08164, 2017. [Online]. Available: http://arxiv.org/abs/1705.08164

[24] G. Ding, Q. Wu, Y.-D. Yao, J. Wang, and Y. Chen, "Kernel-based learning for statistical signal processing in cognitive radio networks: Theoretical foundations, example applications, and future directions," *IEEE Signal Processing Magazine*, vol. 30, no. 4, pp. 126-136, July 2013.

[25] M. R. Vyas, D. K. Patel, and M. Lopez-Benitez, "Artificial neural network based hybrid spectrum sensing scheme for cognitive radio," in *2017 IEEE 28th Annual International Symposium on Personal, Indoor, and Mobile Radio Communications (PIMRC)*, Oct 2017, pp. 1–7.

[26] Y. Tang, Q. Zhang, and W. Lin, "Artificial neural network based spectrum sensing method for cognitive radio," in *2010 6th International Conference on Wireless Communications Networking and Mobile Computing*, 2010, pp. 1–4.

[27] D. Han, G. C. Sobabe, C. Zhang, X. Bai, Z. Wang, S. Liu, and B. Guo, "Spectrum sensing for cognitive radio based on convolution neural network," in *2017 10th International Congress on Image and Signal Processing, BioMedical Engineering and Informatics (CISP-BMEI)*, 2017, pp. 1–6.

[28] Q. Cheng, Z. Shi, D. N. Nguyen, and E. Dutkiewicz, "Deep learning network based spectrum sensing methods for OFDM systems," arXiv: 1807.09414, 2019.

[29] W. M. Lees, A. Wunderlich, P. Jeavons, P. D. Hale, and M. R. Souryal, "Deep learning *classification* of 3.5 GHz band spectrograms with applications to spectrum sensing," aiXiv: 1806.07745, 2018.

[30] Y. LeCun, L. Bottou, Y. Bengio, and P. Haffner, "Gradient-based learning applied to document recognition," *Proceedings of the IEEE*, vol. 86, no. 11, pp. 119-130, 1998.

[31] K. He, X. Zhang, S. Ren, and J. Sun, "Deep residual learning for image recognition," in *Proceedings of IEEE Conference on Computer Vision and Pattern Recognition*, 2016, pp. 770-778.

[32] ARINC618-5, "Air/ground character-oriented protocol, specification," Aeronautical Radio, Inc., August 31, 2000.